\documentclass[draft]{agujournal2019}
\usepackage{apacite}
\usepackage{url} 
\usepackage{xcolor}
\journalname{Water Resources Research}
\begin{document}

\title{A co-spectral budget model links turbulent eddies to suspended sediment concentration in channel flows}
\authors{Shuolin Li\affil{1,2}, Andrew D Bragg\affil{1}, Gabriel Katul\affil{1,2}} 
\affiliation{1}{Department of Civil and Environmental Engineering, Duke University, Durham, NC, USA}
\affiliation{2}{Nicholas School of the Environment, Duke University, Durham, NC, USA}
\correspondingauthor{Shuolin Li}{shuolin.li@duke.edu}

\begin{keypoints}
\item A suspended sediment concentration (SSC) equation for turbulent flows is proposed and tested.
\item The equation is derived from a co-spectral budget that accounts for energy distribution in all eddy sizes.
\item The effects of Reynolds number and a scale-dependent Schmidt number on SSC are explicitly described.
\end{keypoints}

\begin{abstract}
The vertical distribution of suspended sediment concentration (SSC) remains a subject of active research given its relevance to a plethora of problems in hydraulics, hydrology, ecology, and water quality control.   Much of the classical theories developed over the course of 90 years represent the effects of turbulence on suspended sediments (SS) using an effective mixing length or eddy diffusivity without explicitly accounting for the energetics of turbulent eddies across scales.  To address this gap, the turbulent flux of sediments is derived using a co-spectral budget (CSB) model that can be imminently used in SS and other fine particle transport models. The CSB closes the pressure-redistribution effect using a spectral linear Rotta scheme modified to include isotropoziation of production and interactions between turbulent eddies and sediment grains through a modified scale-dependent de-correlation time.  The result is a formulation similar in complexity to the widely used Rouse's equation but with all characteristic scales, Reynolds number, and Schmidt number effects derived from well-established spectral shapes of the vertical velocity and accepted constants from turbulence models. Finally, the proposed CSB model can recover Prandtl's and Rouse's equations under restricted conditions.
\end{abstract}

\section{Introduction}
In his classic treatise on sediment transport, Hans Albert Einstein (HAE) presented a definition of suspended sediments (SS) and the role of turbulence in maintaining suspension as follows \cite{einstein1950bed}: 

{\it "The characteristic definition of a suspended solid particle is that its weight is supported by the surrounding fluid during its entire motion. While being moved by the fluid, the solid particle, which is heavier than the fluid, tends to settle in the surrounding fluid.  If the fluid flow has only horizontal velocities, it is impossible to explain how any sediment particle can be permanently suspended.  Only if the irregular motion of the fluid particles, called turbulence, is introduced can one show that sediment may be permanently suspended}."

This operational definition is now standard in textbooks and research articles alike \cite{dey2014fluvial, green2014review, dey2020fluvial}. Despite some 80 years of research, the dominant factors controlling suspended sediment concentration (SSC) in streams continue to draw interest due to its multiple connections to ecosystem benefits and water quality degradation issues \cite{muste2005two, long2013remote, nazeer2014heavy, dai2016decline, huai2019predicting, huai2020analytical, tsengtwo}. High SSC can intercept photosynthetically active radiation necessary for sustaining submerged aquatic plants in lakes and rivers. The presence of high SSC is also related to eutrophication and corollary water quality issues \cite{yujun2008sediment, kellogg2014use}, clogging of gills of fish and other aquatic organisms, accelerating the denitrification process \cite{liu2013acceleration}.  In certain cases, sediments provide necessary nutrients to aquatic plants and are of primary significance to sustaining nearshore ecosystems such as floodplains and marshes. Their role in element-cycling has been highlighted in several studies \cite{lupker2011rouse, mohtar2020incipient} as well.  Another issue is the connection between SSC and micro/nano-plastics in saline environments.  Recent work has shown that SS can promote polystyrene nano plastics settling in the presence of saline conditions, prompting further interest in SSC distribution in natural waters \cite{li2019interactions}.

Even in the most idealized flow condition with a balance between the gravitational settling flux and the vertical turbulent sediment flux, the description of SSC remains a recalcitrant problem. A model for the turbulent vertical flux is required and is often derived using Reynolds' analogy \cite{dey2014fluvial} where eddies are assumed to transport momentum and SS similarly.  This analogy was the cornerstone of the well-celebrated Rouse's formula \cite{rouse1939analysis} that assumes sediment diffusivity is proportional to eddy viscosity.  Since the early work of O'Brien \cite{obrien1933review}, Prandtl and von K\'arm\'an \cite{vonKarman1934}, these analogies have spawned numerous theories and closure models for the mixing length \cite{vanoni1984fifty,nie2017vertical,bombardelli2009hierarchical,bombardelli2012exchange, dey2014fluvial}. However, these models make no explicit contact with turbulent eddies and their associated kinetic energy distribution in the vertical direction.  It is precisely the scale-wise vertical turbulent kinetic energy component that maintains sediments in suspension \cite{scully2003influence, mazumder2006velocity, dey2014fluvial} as noted by HAE. 

The turbulent vertical flux of SS is directly modeled here from the spectrum of turbulent eddies thereby providing a new perspective on Reynold's analogy, the multiple length scales involved in describing SSC, and the emergence of Reynolds, Rouse, Schmidt, and Stokes numbers when linking eddy viscosity with eddy diffusivity for SS.  The role of the Reynolds number has been introduced in prior studies as a damping correction to the mixing length \cite{van1956turbulent,wallin2000explicit,nezu2004turbulence} whereas the Rouse number is operationally used in the classification of sediment load. The proposed approach uses a co-spectral budget model (CSB) derived from an approximated Navier-Stokes equation in spectral form for the Reynolds stress and SS turbulent flux.  It uses a spectral Rotta scheme modified to include the isotropization of the production term for the pressure decorrelation effect \cite{katul2013co} and a Schmidt number effect similar in form to van Rijin's bulk formulation \cite{rijn1984sediment} for linking the fluid and particle velocity decorrelation time scales, explicitly made here scale-dependent. The newly proposed formulation and a simplified solution derived from it are tested with several published experiments that span a wide range of flow conditions and grain properties (diameter and density). A comparison against the widely-used Rouse formula is featured and discussed.

\section{Theory}
\subsection{Definitions and General Considerations}
As a starting point to review models for SSC profiles in streams, a prismatic rectangular channel with constant width $B$ and bed slope $S_o$ is considered.  The flow is assumed to be steady and uniform with constant water depth $H$ and flow rate $Q$.  For small slopes, a balance between gravitational and frictional forces for a length segment $\Delta x$ along the flow direction $x$ yields
\begin{linenomath*}
\begin{equation} 
\rho (B H \Delta x) g S_o = 2 \tau_s (H \Delta x) + \tau_o (B \Delta x),
\label{eq:forcebalance1}  
\end{equation}
\end{linenomath*}
where $\tau_s$ is the side stress, $\tau_o$ is the bed stress, $g$ is the gravitational acceleration, and $\rho$ is the fluid density.  This expression can be re-arranged as 
\begin{linenomath*}
\begin{equation} 
u_*^2=\frac{\tau_o}{\rho} ={g H S_o}\left(1+ \frac{2 H}{B} \frac{\tau_s}{\tau_o}\right)^{-1},
\label{eq:forcebalance2}  
\end{equation}
\end{linenomath*}
where $u_*$ is the friction (or shear) velocity. For the case where $\tau_s=\tau_o$, $u_*^2=g R_h S_o$ with $R_h=H (1+2 H/B)^{-1}$ being the hydraulic radius.  However, in many SS laboratory experiments, the channel bed is covered with sediments whereas the channel sides remain smooth to permit optical access. This difference in roughness between sides and bed leads to  $\tau_s/\tau_o \ll 1$.  This assumption can be combined with $H/B\le1$ usually selected to minimize secondary circulation to result in $\tau_o/\rho \approx g H S_o $.  This approximation is adopted throughout. Fully turbulent flow conditions are also assumed to prevail so that the bulk Reynolds number $Re_b=U_b H/\nu>500$, where $\nu$ is the kinematic viscosity and $U_b$ is the bulk or depth-averaged velocity given as 
\begin{linenomath*}
\begin{equation} 
\label{eq:Ubulk_def}  
U_b=\frac{Q}{B H} \approx \frac{1}{H} \int_0^H\overline{u}(z)dz,
\end{equation}
\end{linenomath*}
where $\overline{u}(z)$ is the mean velocity at vertical distance $z$ from the channel bed (positive upwards), and overline indicates ensemble-averaging usually determined from time averaging.  For such a flow, the Reynolds-averaged mean continuity equation for SSC in steady and planar homogeneous flow at high $Re_b$ yields \cite{richter2018inertial}
\begin{linenomath*}
\begin{equation} 
\frac{\partial \overline{C}(z)}{\partial t} =0=-\frac{\partial }{\partial z}\left[\overline{w'C'} - w_s\overline{C}-\Phi(z)\right],
\label{eq:fgov}  
\end{equation}
\end{linenomath*}
where $t$ is time, $C=\overline{C}+C'$ is the instantaneous volumetric SSC in the flow, primed quantities are the fluctuating component, $w$ is the instantaneous vertical velocity component with $\overline{w}=0$ (assuming water is of constant $\rho$), $\overline{w'C'}$ is the turbulent vertical flux that requires a closure model, $w_s$ is the terminal velocity of sediment grains, and $\Phi(z)$ arises from particle inertia.  In the regime where particle inertia is weak, to a leading approximation, $\Phi(z)$ is given by \cite{ferry2001fast,richter2018inertial} 
\begin{linenomath*}
\begin{equation} 
\Phi(z)=\tau_p \overline{\left[C \frac{D w'}{Dt} \right]}=\tau_p \overline {C} \frac{\partial \sigma_w^2}{\partial z}+\tau_p \overline{\left[C' \frac{D w'}{Dt} \right]},
\label{eq:Phip}  
\end{equation}
\end{linenomath*}
where $\tau_p=w_s/g$ is a particle time scale, $\sigma_w^2=\overline{w'w'}$ is the vertical velocity variance at $z$ and $D(.)/Dt$ is the material derivative (local and advective) along a fluid particle trajectory.  The $\Phi(z)$ is the sum of a turbophoretic effect that arises due to finite $\partial \sigma_w^2/\partial z$ in inhomogeneous flows such as channels \cite{reeks83,sardina2012wall,johnson20} and a turbulent concentration-vertical acceleration interaction terms.  In equation \ref{eq:fgov}, the overall significance of $\Phi(z)$ at any $z$ depends on a local Stokes number $St(z)=\tau_p/\tau_{K}(z)$ where $\tau_K(z)=[\nu/ \epsilon(z)]^{1/2}$ is the Kolmogorov time scale formed by the local turbulent kinetic energy dissipation rate $\epsilon(z)$ and $\nu$ as reviewed elsewhere \cite{bragg2021mechanisms}. An associated length scale to $\tau_K$ is $\eta=(\nu^3/\epsilon)^{1/4}$, which is the Kolmogorov micro-scale representing eddy sizes impacted by viscous effects at $z$.  Upon defining the Kolmogorov velocity as $v_k=\eta/\tau_K$, the Kolmogorov micro-scale Reynolds number $Re_k=v_k \eta/\nu=1$, meaning that both turbulence and viscous effects are equally important at scales commensurate to $\eta$ \cite{tennekes2018first}.  In the limit $St\to 0$, the particle vertical velocity is given by the sum of the local vertical fluid velocity minus $w_s$, and $\Phi(z)$ can be ignored relative to the turbulent flux at $z$, an assumption routinely invoked in operational models for SSC. To allow for a 'bulk' Stokes number $St_b$ to be formulated, thereby facilitating comparisons across experiments, $\tau_{K,b}=(\nu/ \epsilon_b)^{1/2}$ is proposed where $\epsilon_b$ is the over-all bulk dissipation rate in clear water.  Thermodynamic considerations require that the work per unit mass per unit time to move clear water at $U_b$ is $(g S_o) U_b$.  For steady-state conditions (i.e. turbulent kinetic energy is stationary), this mechanical work produces turbulence that is then dissipated by the action of viscosity leading to an increase in the internal energy of the fluid.  Hence,    
\begin{linenomath*}
\begin{equation} 
\epsilon_b=(g S_o) U_b;~~~ \tau_{K,b}=\sqrt{\frac{\nu}{\epsilon_b}}; ~~~ {\rm and} ~~~ St_b=\left(\frac{w_s}{g}\right)\tau_{K,b}^{-1}.
\label{eq:St_b}  
\end{equation}
\end{linenomath*}
It is assumed that $\Phi$ is small and can be ignored when $St_b\ll 1$ (although, more precisely, $\Phi$ can only be ignored when $\max[St_b,St]\ll1$). Another estimate of bulk Stokes number is $St_+=\tau_p (u_*/H)$ \cite{greimann1999two,greimann2001two}, where $(H/u_*)$ is presumed to represent an outer-layer eddy turnover time.  Noting that $g S_o=u_*^2/H$, the two bulk Stokes numbers can related using $St_b=St_+ (Re_b)^{1/2}$. A critique for using $St_+$ as a bulk Stokes number measure have been discussed elsewhere \cite{greimann1999two,richter2018inertial}.    

\begin{figure} [ht]
\centerline{\includegraphics[angle=0,width=0.63\linewidth]{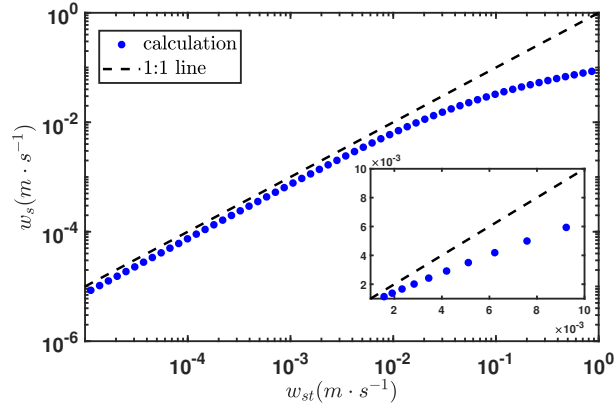}}
\caption{Comparison between the empirical sediment settling velocity $w_s$ used here and the Stokes settling velocity $w_{st}$ for different sediment to fluid density ratios. The one-to-one line is shown. The comparison between $w_s$ and $w_{st}$ for the data sets explored here is also featured as inset. }
\label{fig:ws}
\end{figure}

With regards to the terminal sediment velocity, a simplified expression for $w_s$ that recovers many prior formulae \cite{tan2018rui, huai2020analytical} is used here and is given by \cite{cheng1997simplified}
\begin{linenomath*}
\begin{equation} 
w_s=\frac{\nu}{d_s}\left[\sqrt{25+1.2d_s^2\left(\frac{\rho_s-\rho}{\rho}\frac{g}{\nu^2}\right)^{2/3}}-5\right]^{3/2},
\label{eq:ws}  
\end{equation}
\end{linenomath*}
where $\rho_s$ is the sediment grain density (with $\rho_s/\rho>1$), and $d_s$ is the sediment grain diameter.  This $w_s$ is smaller than the Stokes settling velocity ($w_{st}$)
\begin{linenomath*}
\begin{equation} 
w_{st}=\frac{1}{18} \frac{g}{\nu}\left(\frac{\rho_s-\rho}{\rho}\right) d_s^2 ,
\label{eq:ws_stokes}  
\end{equation}
\end{linenomath*}
\noindent except when $w_{st} d_s/\nu \ll 1$. The comparison between the two settling velocities is shown in Figure \ref{fig:ws} for reference.
Since $w_{st}$ only applies to creeping flow past a sphere, equation \ref{eq:ws} is used as it covers a wider range of $w_s d_s/\nu$. 

The mode of sediment transport is operationally related to $w_s$ and some measure of the strength of turbulence based on bulk flow properties. One such measure is the Rouse number $R$ or 'unit' Rouse number $R^*$ given by
\begin{linenomath*}
\begin{equation}
R^*= \frac{1}{\kappa}\frac{w_s}{u_*}; \quad R=\frac{1}{\beta}R^*;
\label{eq:RN} 
\end{equation}
\end{linenomath*}
where $\kappa=0.41$ is the von K\'arm\'an constant and $\beta=Sc^{-1}$ is an inverse turbulent Schmidt number ($Sc$). The Rouse number is routinely used for classifying sediment load: $R>2.5$ for bedload, $0.8<R<2.5$ for SS, and $R<0.8$ for washload.  To solve for $\overline{C}$, models linking $\overline{w'C'}$ to $\overline{C}$ as well as estimates for $Sc$ (and $\Phi$, though this is ignored here) are required in equation \ref{eq:fgov}, and those models are to be briefly covered. 

\subsection{Conventional Formulations and Revisions}
Conventional approaches (including Rouse and O'Brien) for modeling SSC begin by ignoring $\Phi(z)$ and employing a gradient-diffusion approximation (or some non-Fickian revision to it) given as 
\begin{linenomath*}
\begin{equation} 
\label{eq:gradif} 
\overline{w'C'}=-D_s \frac{d\overline{C}}{dz}, 
\label{eq:graddiff1}
\end{equation}
\end{linenomath*}
where $D_s$ is the sediment turbulent diffusivity. To estimate $D_s(z)$, existing theories approximate $D_s(z)$ by $\nu_t/Sc$ or $\beta \nu_t$, where $\nu_t$ is the turbulent or eddy viscosity ($\nu_t/\nu \gg 1$). When the mixing length hypothesis is further invoked to model $\nu_t$ as a product of a characteristic length and velocity, it yields
\begin{linenomath*}
\begin{equation} 
\nu_t=l_o \left(l_o\,\left|\frac{d\overline{u}}{dz}\right|\right),
\label{eq:nut_l}  
\end{equation}
\end{linenomath*}
where $l_o$ is a generic mixing length to be externally supplied that can vary with $z$.  Dimensional analysis and similarity theory represent 
\begin{linenomath*}
\begin{equation}
\frac{d\overline{u}}{dz} = \frac{\sqrt{-\overline{u'w'}(z)}}{l_o(z)},
\end{equation}
\end{linenomath*}
where $u'$ is the longitudinal velocity fluctuation, and $\overline{u'w'}$ is the momentum turbulent flux at height $z$ that can be estimated from the mean momentum balance using \cite{dey2014fluvial} 
\begin{linenomath*}
\begin{equation} 
\frac{-\overline{u'w'}(z)}{u_*^2}=\left (1-z_n\right),
\label{eq:MMB1}  
\end{equation}
\end{linenomath*}
where $z_n=z/H$ is the normalized water depth.  With this estimate of $\overline{u'w'}(z)$, it follows directly that 
\begin{linenomath*}
\begin{equation}
\frac{d\overline{u}}{dz}=\frac{u_*}{l_o}\left(1-z_n\right)^{1/2}; \nu_t= u_* {l_o}\left(1-z_n\right)^{1/2}; 
D_s=\beta{u_* l_o}\left(1-z_n\right)^{1/2}.
\label{eq:MMB} 
\end{equation}
\end{linenomath*}
These expressions ensure that as $z_n \rightarrow 1$, ${d\overline{u}}/{dz} \rightarrow 0$, $\nu_t \rightarrow 0$, and $D_s\rightarrow 0$.  For $z_n\ll 1$ but $z^+>50$ (i.e. above the buffer layer) where $z^+=z u_*/\nu$ is a normalized distance in wall units \cite{pope2001turbulent} that can also be interpreted as a local Reynolds number ($Re_s$), $l_o$ is constrained by the channel bottom so that $l_o=\kappa z$.  In this case, ${d\overline{u}}/{dz} \approx u_*/(\kappa z)$ and $\overline{u}(z)$ varies logarithmically with $z$,  $\nu_t =\kappa z u_* $, and $D_s=\beta \kappa  z u_*$ (i.e. linear in $z$).  As $z_n \rightarrow 1$, the largest eddies are restricted by $H$ so that $l_o \propto H$ instead of $z$. Combining these two arguments using $l_o=\kappa z (1-z_n)^{1/2}$ yields the quadratic diffusivity profile reported in a number of stream flow studies \cite{fischer2013mixing} and direct numerical simulations (DNS) of stratified atmospheric flows on inclined planes \cite{giometto2017direct}. Assuming $\beta=Sc^{-1}=1$, the SSC profiles associated with the linear and quadratic $D_s(z)$ are
\begin{linenomath*}
\begin{eqnarray} 
\frac{\overline{C}(z_n)}{\overline{C_b}}= \Bigg\{
\begin{array}{l l} 
(\frac{z_n}{z_{n,b}})^{-R^*},                        & \textrm{linear diffusivity}, ~~\textrm{Prandtl's power law} \\  
(\frac{z_n}{1-z_n}\frac{1-z_{n,b}}{z_{n,b}})^{-R^*},   & \textrm{quadratic diffusivity}, ~~\textrm{Rouse's formula}
\end{array},
\label{eq:geS}  
\end{eqnarray}
\end{linenomath*}
where $\overline{C_b}$ is a reference concentration at height $z_{n,b}=z_b/H$ and $R^*=R$ when setting $\beta=1$. The $R^*$ in equation \ref{eq:geS} is commonly replaced by a fitted $R$ (or $\beta$ is no longer unity) as discussed elsewhere \cite{muste2005two, dey2014fluvial}. The analysis using fitted $R$ is termed here as 'fitted' Rouse's formula.  Other models for $l_o$ have been introduced but only two are singled out for illustrating differences in approaches to adjusting conventional formulations (usually for $\kappa z$): (i) $l_o=\kappa z V_n(z_n)$, where $V_n=1-\exp(-z^+/26)$ (labeled as the van Driest damping function); (ii) $l_o=\kappa z (1-z_n)^{m_1}$, where
\begin{linenomath*}
\begin{equation}
m_1=\frac{1}{2}\left[1+a_e \left(\frac{\overline{C}}{C_R}\right)\right],
\label{eq:loCz} 
\end{equation}
\end{linenomath*}
$C_R$ is some reference concentration and $a_e$ is an empirical coefficient \cite{umeyaina1992vertical,mazumder2006velocity,castro2012karman}.  In the second case, the mixing length is assumed to vary with SSC and recovers $l_o=\kappa z (1-z_n)^{1/2}$ only for clear water. However, in the presence of sediments, $m_1$ varies with $z_n$ (and $R$).  In the first case, deviations from a linear mixing length is made to dependent on $z^+$ (instead of $H$), which is appropriate in the viscous and buffer regions of smooth boundary layers. Another revision to equation \ref{eq:graddiff1} is to re-cast turbulent transport in fractional derivatives to emphasize its non-Fickian aspect \cite{nie2017vertical}. In this approach, the fractional order becomes a parameter that must be determined from experiments depending on how SS trajectories deviate from Brownian trajectories \cite{sun2020review}. In practice, the order of the fractional derivative is set as a 'free' parameter and must implicitly include the $Sc$ effect. This approach is not pursued further here. 

\subsection{Turbulent Stress and SS Flux Budgets}
Simplified turbulent stress and SS flux budgets are now considered.  For a stationary and planar homogeneous flow in the absence of subsidence ($\overline{w}=0$), these budgets reduce to 
\begin{linenomath*}
\begin{eqnarray}
\frac{\partial{\overline{w'u'}}}{\partial t} =0 &=& -\overline{w'w'}\frac{\partial \overline{u}}{\partial z}-\frac{\partial{\overline{w'w'u'}}}{\partial z}+\overline{p'\frac{\partial u'}{\partial z}}-\epsilon_{wu}, ~~\\ \nonumber
\frac{\partial{\overline{w'C'}}}{\partial t}=0&=&-\overline{w'w'}\frac{\partial \overline{C}}{\partial z}-\frac{\partial{\overline{w'w'C'}}}{\partial z}+\overline{p'\frac{\partial C'}{\partial z}}-\epsilon_{wc} -w_s\overline{ \left(w' \frac{\partial C'}{\partial z} \right)},
\label{eq:stressfluxbudget} 
\end{eqnarray}
\end{linenomath*}
where $p'$ is the turbulent pressure, $\epsilon_{wu}$ and $\epsilon_{wc}$ are molecular destruction terms assumed to be small when compared to the pressure-decorrelation terms at high Reynolds numbers \cite{katul2013co}.  The turbulence- particle interaction term requires closure that may be achieved by commencing with a local decomposition given by,
\begin{linenomath*}
\begin{eqnarray}
w_s\overline{\left(w' \frac{\partial C'}{\partial z} \right)}
=w_s\left[\left(\frac{\partial \overline{w' C'}}{\partial z} \right)-
\left(\overline{C'\frac{\partial w'}{\partial z}} \right)\right].
\label{eq:stressfluxbudget1} 
\end{eqnarray}
\end{linenomath*}
When assuming $\Phi(z)=0$ in equation \ref{eq:fgov} (i.e. no particle inertia), $\overline{w'C'}=w_s \overline{C}$ thereby allowing one of the two terms in the difference shown in equation \ref{eq:stressfluxbudget1} to be linked to variables that are explicitly modeled. The other term (i.e. $\overline{C'{\partial w'}/{\partial z}}$) still necessitates a closure.  A heuristic model that maintains maximum simplicity is to set
\begin{linenomath*}
\begin{equation}
\overline{C'\frac{\partial w'}{\partial z}}=b_1 \frac{{\partial \overline{w' C'}}}{{\partial z}},
\label{closure_1_RANS}
\end{equation}
\end{linenomath*}
where $b_1$ is a positive or a negative constant.  Upon setting  $\overline{w'C'}=w_s \overline{C}$, this heuristic closure model yields \cite{huang2014particle},
\begin{linenomath*}
\begin{eqnarray}
w_s\overline{\left(w' \frac{\partial C'}{\partial z} \right)}
= w_s\left[\left(\frac{\partial w_s \overline{C}}{\partial z} \right)-b_1 \left(\frac{\partial w_s \overline{C}}{\partial z} \right)
\right]= \alpha' w_s^2 \frac{\partial \overline{C}}{\partial z},
\label{eq:stressfluxbudget3} 
\end{eqnarray}
\end{linenomath*}
where $\alpha'=1-b_1$ is a constant.  When $|b_1|\ll 1$, then $\alpha'=1$ and 
\begin{linenomath*}
\begin{eqnarray}
w_s\overline{\left(w' \frac{\partial C'}{\partial z} \right)}
=w_s^2\frac{\partial \overline{C} }{\partial z} .
\label{eq:stressfluxbudget4} 
\end{eqnarray}
\end{linenomath*}
Whether $b_1$ or $\alpha'$ are strictly closure constants independent of sediment and/or flow conditions cannot be a priori ascertained.  To do so requires another scaling analysis based on different assumptions and approximations.   In this proposed scaling analysis, $C'$ is assumed to vary with a turbulent quantity such as $\sigma_c$, and $w'$ to vary with $\sigma_w$.  Hence, 
\begin{linenomath*}
\begin{eqnarray}
\overline{\left(C' \frac{\partial w'}{\partial z} \right)}
=A_F \left[\sigma_c(z)\right] \frac{\partial \sigma_w}{\partial z} =A_F \left[\frac{\overline{w'C}}{u_*}F_1 \left(z_n\right)\right]\frac{\partial \sigma_w}{\partial z}, 
\label{eq:close1} 
\end{eqnarray}
\end{linenomath*}
where $A_F$ is a flux-variance \cite{albertson1995sensible} similarity constant that can be positive or negative depending on the sign of the correlation coefficient between $C'$ and $\partial w'/\partial z$, and $F_1(z_n)$ is an unknown dimensionless function describing the sediment concentration variance with $z_n$ above and beyond the $\overline{w'C'}$ variations with $z_n$.  Since the goal is to determine the minimum governing variables impacting $b_1$ or $\alpha'$ while assuming $b_1$ is independent of $z_n$, equations \ref{eq:close1} and \ref{closure_1_RANS} can be equated to yield 
\begin{linenomath*}
\begin{eqnarray}
A_F \left[\frac{\overline{w'C'}(z_n)}{u_*}F_1 \left(z_n\right)\right]\frac{\partial \sigma_w}{\partial z}=b_1 \frac{{\partial \overline{w' C'}(z_n)}}{{\partial z}}. 
\label{eq:close1a} 
\end{eqnarray}
\end{linenomath*}
Re-arranging to infer $b_1$ results in
\begin{linenomath*}
\begin{eqnarray}
b_1= A_F \left[\overline{w'C'}\left(\frac{\partial \overline{w' C'}}{\partial z}\right)^{-1}\right] \left[\frac{1}{u_*}F_1 \left(z_n\right)\frac{\partial \sigma_w}{\partial z}\right].
\label{eq:close1b} 
\end{eqnarray}
\end{linenomath*}
With the assumption that $b_1$ is not dependent on $z_n$, additional order of magnitude arguments must now be invoked to assess the sediment/flow variables that impact its magnitude: (i) $\partial \sigma_w/\partial z\sim -u_*/H$ (likely valid except near the channel bottom), (ii) $F_1$ is roughly a constant,  (iii) $\partial \overline{w' C'}/{\partial z} = w_s \partial \overline{C}/\partial z$, and (iv) $\overline{w'C'}/(\partial \overline{C}/\partial z)\sim - D_{s,avg}$ where $D_{s,avg}=(1/H)\int_0^H D_s(z)dz \sim u_* H$.  Inserting these order of magnitude arguments into equation \ref{eq:close1b} result in
\begin{linenomath*}
\begin{eqnarray}
b_1 \sim \mathrm{sgn}(A_f) \frac{u_* H}{w_s} \left[\frac{1}{u_*} \frac{u_*}{H}\right] \sim \mathrm{sgn}(A_f) \frac{u_*}{w_s}.
\label{eq:close1c} 
\end{eqnarray}
\end{linenomath*}
Equation \ref{eq:stressfluxbudget1} is used to suggest a pragmatic closure in equation \ref{eq:stressfluxbudget3} that applies to only one of two terms, and this one term itself is only one term in the overall flux budget.  Given the interplay between these multiple terms, the overall model results for $\overline{C}$ may be robust to uncertainties in this closure vis-a-vis externally imposing $Sc$ or $\beta^{-1}$ directly on the eddy diffusivity as common in prior models.     

Upon ignoring the flux transport terms (triple moments), and closing the pressure decorrelation terms using a linear Rotta scheme that accounts for the isotropization of the production yields
\begin{linenomath*}
\begin{eqnarray}
-(1-C_I) \sigma_{w}^2\frac{\partial \overline{u}}{\partial z}-A_R \frac{\overline{w'u'}}{\tau}=0 , \quad
\left[ -(1-C_I) - 
\alpha' \frac{ w_s^2}{\sigma_{w}^2} \right] 
\sigma_{w}^2\frac{\partial \overline{C}}{\partial z}-A_R \frac{\overline{w'C'}}{\tau}=0,
\label{eq:stressfluxbudget2} 
\end{eqnarray}
\end{linenomath*}
where $\tau$ is a turbulent relaxation time scale, $C_I=3/5$ is the isotropization of the production constant determined from rapid distortion theory \cite{pope2001turbulent}, and $A_R=1.8$ \cite{katul2013co,katul2014cospectral} is the Rotta constant assumed to be the same for momentum and SS.  It directly follows from these simplified budgets that a model of maximum simplicity for $Sc$ may be derived as
\begin{linenomath*}
\begin{equation} 
Sc^{-1}(z_n)=\frac{D_s}{\nu_t}=  1 + \alpha \left(\frac{ w_s}{\sigma_{w}}\right)^2,
\label{eq:wc1}  
\end{equation}
\end{linenomath*}
where $\alpha=\alpha'/(1-C_I)$, though $\alpha'$ or $b_1$ can vary themselves with $u_*/w_s$ as noted earlier.  It is necessary to point out that when $\alpha\geq 0$, equation \ref{eq:wc1} is opposite to what is predicted by the so-called 'crossing-trajectories' effect for heavy particles settling in a turbulent flow. The crossing trajectories arise when particle trajectories cross trajectories of fluid elements under the influence of gravity. This effect invariably forces particles to move from a region of highly correlated flow to another less correlated region \cite{wells1983effects}. In this manner, particles lose velocity correlation more rapidly than the corresponding fluid points and thus must disperse less. Thus, the crossing trajectories effect requires $Sc>1$ \cite{csanady1963turbulent, duman2016dissipation}. 

\subsection{The Co-spectral Budget Model}
The models so far make no explicit contact with the phenomenon they perpetrate to represent: turbulent eddies and their energy distribution. The proposed approach here uses a co-spectral budget model (CSB) to achieve such a link.  The CSB is derived from an approximated Navier-Stokes equation in a spectral form that links turbulent eddies of different sizes to $\overline{w'C'}$.  The CSB derivation commences by noting that $\overline{w'C'}$ and $\overline{u'w'}$ both satisfy the normalizing properties, 
\begin{linenomath*}
\begin{equation} 
-\overline{w'C'} =\int_{0}^{\infty}\phi_{wc}(k)dk, ~~ -\overline{u'w'} =\int_{0}^{\infty}\phi_{wu}(k)dk,
\label{eq:wc}  
\end{equation}
\end{linenomath*}
where $\phi_{wc}(k)$ and $\phi_{wu}(k)$ are the co-spectral density functions of the turbulent vertical velocity-turbulent sediment concentration and turbulent vertical-longitudinal velocities, respectively, and $k$ is the wavenumber or inverse eddy size. The co-spectral budgets associated with equation \ref{eq:stressfluxbudget} have been derived elsewhere and simplify to \cite{bos2004behavior, cava2012scaling,katul2013co,katul2014cospectral},
\begin{linenomath*}
\begin{eqnarray} 
\frac{\partial}{\partial t} \phi_{wu}(k)=0 =P_{wu}(k)+T_{wu}(k)+\pi_{wu}(k)-2\nu k^2\phi_{wu}(k),\\
\frac{\partial}{\partial t} \phi_{wc}(k)=0 =P_{wc}(k)+T_{wc}(k)+\pi_{wc}(k)-\nu(1+Sc_m^{-1})k^2\phi_{wc}(k),
\label{eq:vscg}  
\end{eqnarray}
\end{linenomath*}
where $P_{wu}(k)=({d\overline{u}}/{dz}) E_{ww}(k)$ and $P_{wc}(k)=({d\overline{C}}/{dz}) E_{ww}(k)$ are the stress and flux production terms at $k$, $E_{ww}(k)$ is the vertical velocity spectrum satisfying the normalizing relation $\sigma_w^2 = \int_0^{\infty} E_{ww}(k)dk$, $T_{wu}(k)$ and $T_{wc}(k)$ are turbulent transfer terms, $\pi_{wu}(k)$ and $\pi_{wc}(k)$ are pressure-velocity and pressure-scalar decorrelation terms, and $Sc_m$ is the molecular Schmidt number (not related to $Sc$). Invoking a spectral-based Rotta model that includes the isotropization of the production as before, the pressure-scalar co-variance in k-space can be modeled as
\begin{linenomath*}
\begin{eqnarray}
\pi_{wu}(k)=-A_R\frac{1}{t_{ww}(k)}\phi_{wu}(k)-C_IP_{wu}(k)
,\quad
\pi_{wc}(k)=-A_R\frac{1}{t_r(k)}\phi_{wc}(k)-C_IP_{wc}(k),
\label{eq:pi} 
\end{eqnarray}
\end{linenomath*}
where $A_R\approx1.8$ and $C_I=3/5$ are as before, $t_{ww}(k)$ and $t_r(k)$ are the decorrelation time-scale of the turbulent stress and particle concentration.  A model of maximum simplicity is to assume that these two wavenumber dependent time scales are related using a wavenumber-dependent $Sc(k)$ given by,
\begin{linenomath*}
\begin{equation}
t_r(k)={t_{ww}(k)}{Sc^{-1}(k)}, ~~\textrm{with}~
{Sc^{-1}(k)}={1+\alpha{(w_s\,k\,t_{wc})^2 }},
\label{eq:tm} 
\end{equation}
\end{linenomath*}
where $t_{wc}=\min(t_{ww}, f_o~t_{K,b})$ with $f_o$ being a constant (a plausibility argument to such $t_{wc}(k)$ representation is discussed later), $Sc$ is modeled in analogy to equation \ref{eq:wc1} albeit in a spectral form e.g. the local characteristic turbulent velocity is estimated by $(kt_{wc})^{-1}$ using a one-way coupling approach \cite{elghobashi1994predicting}, $t_{ww}(k)\propto \epsilon^{-1/3} k^{-2/3}$ is interpreted as a characteristic time scale derived from dimensional analysis assuming $\epsilon$ is the conserved quantity across the energy cascade of $E_{ww}(k)$, and $\epsilon$ is the turbulent kinetic energy dissipation rate.  One plausible choice for the proportionality constant is $C_o^{-1/2}$ so as to recover a Kolmogorov time scale in the inertial subrange, where $C_o=0.65$ is the Kolmogorov constant for the vertical velocity component. 

\begin{figure} [ht]
\centerline{\includegraphics[angle=0,width=0.72\linewidth]{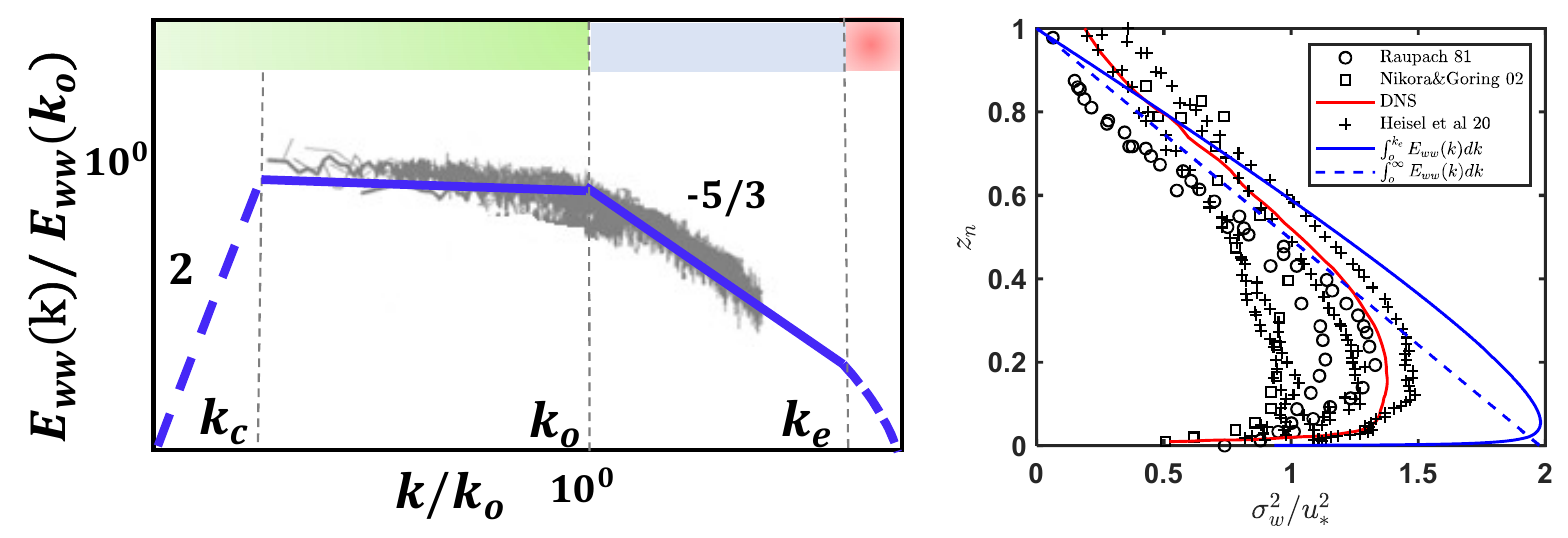}}
\caption{Left: A typical $E_{ww}(k)$ at $z_n$ from the channel bottom. The very low wavenumber range are assumed to follow the Saffman spectrum ($E_{ww}(k)$ $\propto$ $k^2$) until $k_c=1/H$. The Saffman spectrum is then connected using a flat transition (i.e. wall effects introduce energy splashing) to the inertial subrange at $k_o$ $\propto$ $1/z$ where $E_{ww}(k)$ $\propto$ $k^{-5/3}$. The black curves are extracted from measurements \cite{nikora2002fluctuations} with different flow conditions using ADV and do not resolve the viscous dissipation range in the vicinity of $k_e=1/\eta$ or the presumed Saffman spectrum. Right: The $\sigma_w^2/u_*^2$ profile modeled from scale-wise integration of $E_{ww}(k)$ and its simplified form (i.e. ignoring the Saffman contribution and extending the inertial subrange indefinitely to fine scales). The measured $\sigma_w^2/u_*^2$ profiles are from experiments described elsewhere \cite{Raupach81, nikora2002fluctuations, heisel2020velocity}.  They include field experiments and wind-tunnel experiments over a wide range of roughness types and Reynolds number conditions.  The direct numerical simulations (DNS) for a smooth channel (red) are also included for comparisons \cite{heisel2020velocity}. }
\label{fig:etke}
\end{figure}

For scalewise integration, it is also necessary to maintain a bounded $t_r(k)$ as $k \rightarrow 0$ for any $z_n$.  We set $t_r(k)=t_r(k_c)$ when $k<k_c$, where $k_c$ is the smallest inverse length scale where $E_{ww}(k)$ increases with increasing $k$. The viscous-destruction terms are negligible when compared to the Rotta terms for $k\eta\ll1$.  Since $T_{wu}(k)$ and $T_{wc}(k)$ do not contribute to the net production or destruction of $\phi_{wu}(k)$ and $\phi_{wc}(k)$ but only redistribute them across scales (i.e. $\int_0^{\infty}T_{wu}(k)dk=0$, and $\int_0^{\infty}T_{wc}(k)dk=0$), they are ignored for simplicity \cite{bonetti2017manning}. Adopting these simplifications, 
\begin{linenomath*}
\begin{eqnarray}
\phi_{uw}(k) =\left(\frac{1-C_I}{A_R}\right)\frac{d\overline{u}}{dz}\left[E_{ww}(k) t_{ww}(k)\right]
,\quad
\phi_{wc}(k) =\left(\frac{1-C_I}{A_R}\right)\frac{d\overline{C}}{dz}\left[E_{ww}(k) t_r(k)\right].
\label{eq:cvsc0}  
\end{eqnarray}
\end{linenomath*}
To integrate these equations across $k$ and derive turbulent shear stress and sediment flux at any height $z_n$, an expression for $E_{ww}(k)$ is required. A model for $E_{ww}(k)$ that captures known spectral features at an arbitrary $z_n$ is shown in Figure \ref{fig:etke}.

The $E_{ww}(k)$ is now piece-wise approximated as
\begin{linenomath*}
\begin{equation} 
\label{eq:ewweq} 
E_{ww}(k)= \left\{ \begin{array}{l l} 
E_{kol}(k_o)k_c^{-2}k^2,   & \mathrm{if} ~0 \leq k \leq k_c\\
E_{kol}(k_o),              & \mathrm{if} ~k_c \leq k \leq  k_o\\
E_{kol}(k),                & \mathrm{if} ~k_o \leq k \leq  k_e
\end{array} \right.,
\end{equation}
\end{linenomath*}
where $k_c = H^{-1}$, $k_o=(\kappa z)^{-1}$ and $k_e=\eta^{-1}$ are three characteristic wavenumbers that mark the key transitions in $E_{ww}(k)$ between $H$ and the characteristic eddy scales bounding the inertial subrange \cite{bonetti2017manning, katul2013co,li2019cospectral, ayet2020scaling}, and $E_{kol}(k)=C_o\epsilon(z)^{2/3}k^{-5/3}$ is the Kolmogorov spectrum.  In the case of $E_{kol}(k)$, the transfer of energy across scales shapes the energy cascade and is necessary for obtaining the $k^{-5/3}$ scaling.  The transfer of stress across scales, as given by $T_{wu}(k)$, was ignored in the CSB model here.  The inclusion of the transfer term in the energy cascade (indirectly specified by $E_{ww}(k)$) but not in the CSB may appear paradoxical.  This is not so as the role and significance of the transfer terms are quite different when analyzing scale-wise energy and stress budgets \cite{bos2004behavior}. In the inertial subrange where $E_{kol}(k)\sim k^{-5/3}$, a $\phi_{uw}(k) \sim k^{-7/3}$ has also been reported and confirmed in numerous boundary layer experiments and simulations of wall-bounded flows \cite{pope2001turbulent}.  A balance between production and dissipation terms in the CSB model leads to a $\phi_{uw}(k) \sim (d\overline{u}/dz) E_{ww}(k) t_{ww}(k)$, which recovers the $[(d\overline{u}/dz) \epsilon^{1/3}] k^{-7/3}$ scaling in the inertial subrange.  Inclusion of $T_{wu}(k)$ necessarily leads to  $\phi_{uw}(k)$ that must deviate from a $k^{-7/3}$ scaling in the inertial subrange as discussed elsewhere \cite{li2015revisiting}.  Moreover, the constants emerging from a production balancing dissipation in the scale-wise CSB model for the inertial subrange, $[(1-C_I)/A_R] C_o^{1/2}=0.18$, does recover the accepted co-spectral similarity constant whose numerical value was determined at $0.15-0.16$ from wind tunnel studies, atmospheric surface layer studies, and DNS \cite{katul2013co}.  For these reasons (i.e. $T_{wu}(k)$ ignored within the inertial subrange) and because $\int_0^{\infty} T_{wu}(k)dk=0$, $T_{wu}(k)$ is ignored at all $k$.  This assumption is also compatible with ignoring the triple moments in equations \ref{eq:stressfluxbudget}.       

The only remaining term needed to describe the magnitude of $E_{ww}(k)$ at all $k$ is $\epsilon(z)$.  A model of maximum simplicity is to relate $\epsilon(z)$ to the mechanical production $P_{wu}(z)$ of the turbulent kinetic energy budget using \cite{pope2001turbulent} 
\begin{linenomath*}
\begin{equation} 
\epsilon (z)=\frac{P_{wu}(z)}{\phi(z_n)}
=\phi^{-1}(z_n)\left(-\overline{u'w'}\frac{d\overline{u}}{dz}\right)
=\phi^{-1}(z_n)u_*^2\left(1-\frac{z}{H}\right)\frac{d\overline{u}}{dz},
\label{eq:dissipation} 
\end{equation}
\end{linenomath*}
where $\phi(z_n)$ is a modification function to account for the imbalance between the local mechanical production and local dissipation terms in the turbulent kinetic energy budget.  For stationary and planar-homogeneous flow conditions without any mean vertical advection and in the absence of any transport terms, $\epsilon(z)\approx P_{wu}(z)$ and $\phi(z_n)\approx1$. While this estimate may be acceptable in the log-region describing $\overline{u}(z)$, deviations near the channel bottom ($\phi(z_n)>1$) and near the water surface ($\phi(z_n)<1$) are expected.  Hence, $\phi(z_n)$ must be viewed as a depth-dependent function \cite{kim1987turbulence, pope2001turbulent} though its variation from unity is not considered here to maintain maximum simplicity. A plausibility argument for ignoring its variation from unity is that $\overline{w'C'} \propto \left[\phi(z_n)\right]^{-1/3}$ (shown later), which makes the SSC calculations less sensitive to $\phi(z_n)$ deviations from unity. This point is considered later in the context of modeling $\sigma_{w}^2(z_n)$ based on the assumed $E_{ww}(k)$ shape.

Returning to the choice of $t_{wc}=\min(t_{ww}, f_o~t_{K,b})$ and the choice $f_o$, as $z_n\rightarrow1$, $\overline{w'u'}\rightarrow0$, $P_{wu}(z_n)\rightarrow0$, and thus $\epsilon\rightarrow$ (i.e. no turbulence) near the free water surface.  With  $\epsilon\rightarrow$, $t_{ww}(k)\rightarrow\infty$ (along with $\tau_k\rightarrow\infty$ and $\eta_k\rightarrow\infty$).  That $t_{ww}(k)\rightarrow\infty$ is not problematic for the closure scheme of $\pi_{wu}(k)$ and $\pi_{wc}(k)$ as those terms are expected to decay near the free water surface and this decay remains compatible with $t_{ww}(k)\rightarrow\infty$.  The problem of $\epsilon\rightarrow0$ arises in maintaining a finite $Sc^{-1}(k)$ dominated by turbulent processes thereby necessitating a finite $\epsilon$ in the calculation of $Sc^{-1}(k)$ that cannot be readily inferred from $P_{wu}(z_n)$.  To ensure that the particle interaction time scale $t_{wc}$ remains bounded in $Sc^{-1}(k)$, an adhoc minimal value of $\epsilon$, set to be $0.1\% ~ \epsilon_b$, is proposed.  This choice of minimal $\epsilon_b$ prevents $\epsilon\rightarrow0$ as $z_n\rightarrow1$ in the $Sc(k)$ formulation only.  This minimal threshold set to ensure a finite $\epsilon$ in $Sc(k)$ (mainly near the free water surface) leads to $f_o=\sqrt{1000}\approx 31$.

\section{Results and Discussion}

\subsection{Co-spectral Budget Model}
By scale-wise integrating $\phi_{uw}(k)$ and using $u_*^2(1-z_n)=\int_o^{k_e}\phi_{uw}(k)$ dk, the velocity gradient $d\overline{u}/dz$ at $z$ is obtained as 
\begin{linenomath*}
\begin{eqnarray}
\frac{d\overline{u}}{dz}=A_{\pi}^{-3/4}\phi^{1/4}(z_n)\left(1-\frac{z}{H}\right)^{1/2}
\left[
\frac{15}{4}-\frac{8}{3}\left(\frac{k_c}{k_o}\right)^{1/3}-\frac{3}{4}\left(\frac{k_o}{k_e}\right)^{4/3}
\right]^{-3/4}\left(k_o u_*\right),
\label{eq:sscb3}
\end{eqnarray}
\end{linenomath*}
where $A_{\pi}=({1-C_I})\sqrt{C_o}/{A_R}\approx0.18$, and the vertical velocity variance can be derived by scale-wise integrating $E_{ww}(k)$ as,
\begin{linenomath*}
\begin{eqnarray}
\frac{\sigma_{w}^2}{u_*^2}=
\frac{5}{2}C_oA_{\pi}^{-1/2}\phi^{-1/2}(z_n)
\left[1-\frac{4}{15}\frac{k_c}{k_o}
-\frac{3}{5}\left(\frac{k_e}{k_o}\right)^{-2/3}
\right]
\left[\frac{15}{4}-\frac{8}{3}\left(\frac{k_c}{k_o}\right)^{1/3}
-\frac{3}{4}\left(\frac{k_e}{k_o}\right)^{-4/3}
\right]^{-1/2}
(1-z_n).
\label{eq:sigw}
\end{eqnarray}
\end{linenomath*}
Likewise, the SSC turbulent flux is solved as
\begin{linenomath*}
\begin{eqnarray}
-\overline{w'C'}=A_{\pi}\phi^{-1/3}(z_n)\Omega(z)u_*^{2/3}
\left[\left(1-\frac{z}{H}\right)\frac{d\overline{u}}{dz}\right]^{1/3}\frac{d\overline{C}}{dz}
=-w_s\overline{C},
\label{eq:sscb1}
\end{eqnarray}
\end{linenomath*}
with $\Omega(z_n)$ given by
\begin{linenomath*}
\begin{eqnarray}
\Omega(z_n)=\int_{0}^{k_c}Sc^{-1}(k_c)k_c^{-8/3}k_o^{-5/3}k^{2}dk
+\int_{k_c}^{k_o}k_o^{-5/3}Sc^{-1}(k)k^{-2/3}dk+
\int_{k_o}^{k_e}Sc^{-1}(k)k^{-7/3}dk.
\label{eq:sscb11}
\end{eqnarray}
\end{linenomath*}
Therefore, the turbulent Schmidt number $Sc(z_n)$ can be determined from the CSB model as
\begin{linenomath*}
\begin{eqnarray}
Sc(z_n)=\frac{\nu_t}{D_s}=\Omega^{-1}(z_n)\left[
\frac{15}{4}-\frac{8}{3}\left(\frac{k_c}{k_o}\right)^{1/3}-\frac{3}{4}\left(\frac{k_o}{k_e}\right)^{4/3}
\right]k_o^{-4/3}.
\label{eq:sc}
\end{eqnarray}
\end{linenomath*}
Because the determination of $k_e=1/\eta$ (where $\eta=(\nu^3/\epsilon)^{1/4}$) requires an estimate of $\epsilon(z_n)=P_{uw}(z_n)$ and thus an estimate of $d\overline{u}/dz$, an iterative scheme is needed to determine $d\overline{u}/dz$ and $k_e$ at every $z_n$ from equation \ref{eq:sscb3}. Once determined, the $E_{ww}(k)$, $Sc(z_n)$, $\overline{w'C'}$ and the subsequent SSC profile can be computed at each $z_n$ by solving equations \ref{eq:sigw}, \ref{eq:sscb1}, and \ref{eq:sc} for $\sigma_w^2$, $\overline{w'C'}$ and $Sc$. Since there is no analytical solution to this system, a numerical integration using a 3rd-order Adams–Bashforth method is employed.

Before proceeding to the analysis of SSC, an assessment of the assumed shape of $E_{ww}(k)$, its transition wavenumbers, as well as the consequence of the assumption of $\phi(z_n)\approx 1$ is conducted in Figure \ref{fig:etke}.  The predicted $\sigma_w^2/u_*^2$ and its simplified version using $E^1_{ww}(k)$ without the Saffman spectrum and assuming $k_e\rightarrow\infty$ are compared against two sets of experiments: (i) wind tunnel experiments conducted over a wide range of surface roughness types \cite{Raupach81} and (ii) field experiments \cite{nikora2002fluctuations} of the sediment flow in the Balmoral Irrigation Canal (New Zealand).  The wind-tunnel experiments used a hot-wire probe whereas the field experiments used acoustic Doppler velocity (ADV) measurements that do not resolve the viscous dissipation regime.  As expected, the predicted $\sigma_w^2/u_*^2$ here exceeds the measurements because the spectral shapes assumed in $E_{ww}(k)$ account for a much broader range of eddy sizes than the experiments interrogate. Specifically, the Saffman and dissipation ranges are not resolved by the flume experiments whereas the wind tunnel experiments resolve a limited dissipation range but are not conducted over a sufficiently long enough sampling period to cover the Saffman spectrum.  Nonetheless, the model recovers key features of the $(\sigma_w/u_*)^2$ profile: a rapid increase with $z_n$ near the surface, a peak at $(\sigma_w/u_*)^2=1.9$, and a quasi-linear decline as $z_n\rightarrow1$.  The peak $(\sigma_w/u_*)^2=1.9$ is compatible with near-neutral atmospheric surface layer measurements ($=1.8$) where lateral confinements of the flow are absent (unlike flumes and wind tunnels) and where $H/\eta$ far exceeds those obtained in laboratory studies. 

\begin{figure} [ht]
\centerline{\includegraphics[angle=0,width=1.2\linewidth]{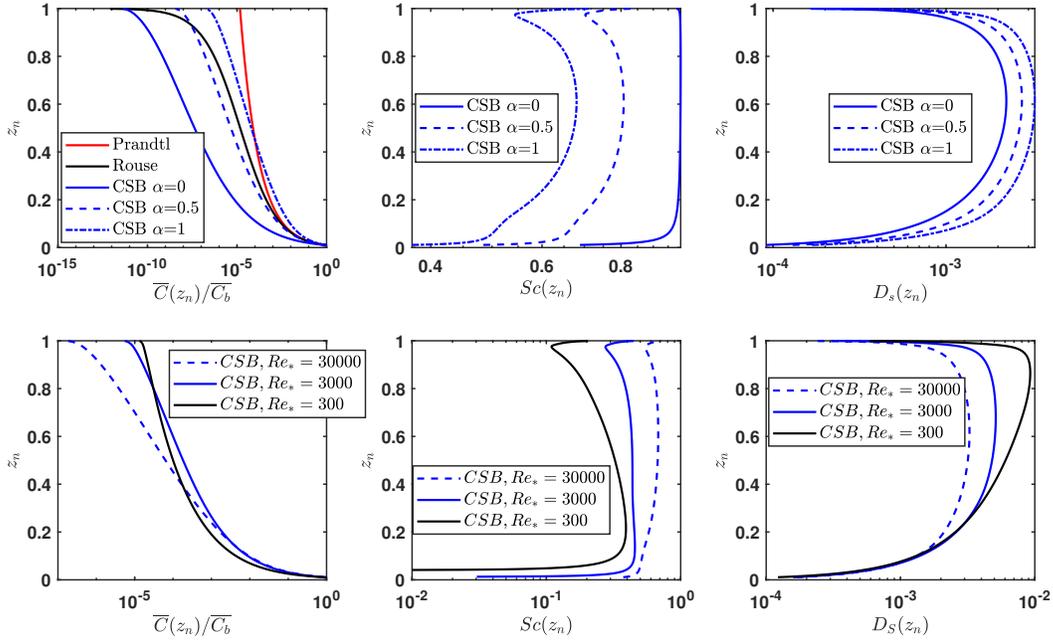}}
\caption{The predicted SSC, $Sc$, and $D_s$ profiles based on the CSB model when setting $d_s$=1 mm, $\rho_s$=1.2 g cm$^{-3}$, $u_*$=3 cm  s$^{-1}$, and $U_b/u_*$=$10$. The reference position is at $z_{n,b}$=$0.01$. Different $\alpha$ values and Reynolds numbers ($Re_*$=$u_*H/\nu$) are featured to illustrate overall sensitivity of the normalized SSC profile to these parameters.  The Prandtl and Rouse model predictions of SSC are shown for reference in the top-left panel.  The Reynolds number is varied by alerting $\nu$.}
\label{fig:SSC_CSB}
\end{figure}

Now the comparisons of the CSB results with $\alpha$ temporarily set as a 'free' parameter with (i) Prandtl's power law solution and (ii) Rouse's formula are shown in Figure \ref{fig:SSC_CSB}.
The computed SSC and $Sc$ profiles are also presented when the flow conditions and sediment properties are externally supplied. For Prandtl's power-law and Rouse's formula, the bulk Schmidt number was set to unity. However, the CSB model allows for a depth-dependent $Sc(z_n)$, which is set by $\alpha$. When $\alpha=0$, $Sc(z_n)=1$ in the entire channel, consistent with equation \ref{eq:wc1}. When $\alpha>0$, $Sc(z_n)$ varies with depth and is generally greater in the near-bed region and becomes smaller with increasing $z_n$.  However, because of the imposition of a finite $\epsilon$ near the water surface ($=0.001\epsilon_b$), $Sc(z_n)$ increases back to near unity when $z_n \rightarrow 1$. Rouse's equation and CSB models exhibit different behavior near the water surface.  Rouse's equation yields a zero-concentration at $z_n=1$ whereas the CSB model does not. One advantage to the CSB approach is its ability to resolve the dependence of $\overline{C}/\overline{C_b}$ on Reynolds number.   Using different $\nu$, variations in $Re_*=u_*H/\nu$ can be generated and their effects on CSB model predictions tracked.  Recall that $H/\eta_b$ (modeled in the CSB) scales as $Re_*^{3/4}$, and the effects of this scale separation on the shape of the vertical velocity spectrum, sediment flux co-spectrum, and the resulting $\overline{C}/\overline{C_b}$ profiles are explicitly determined. The effects of $\alpha$ are much more significant than the effects of $Re_*$, which is heuristically supportive for using Direct Numerical Simulation runs (lower $Re_*$) to further explore the CSB approach. As earlier noted, the implications of setting $t_{wc}=\min(t_{ww}, f_o ~t_{K,b})$ with $f_o=\sqrt{1000}$ are most visible on the $Sc(z_n)$ profile near the free water interface. Altering $f_o$ primarily modifies the thickness of the region near the water interface impacted by the imposed finite $t_{wc}$ (or finite $\epsilon$ in the $Sc(k)$ determination).  However, the CSB model itself is not expected to be valid in this zone as the assumed shape of $E_{ww}(k)$ is not realistic, the flux transport terms can be finite, and turbo-phoretic effects may also be large in this vicinity.  In sum, predictions from the CSB model near the free water surface must be treated with skepticism and caution. 

\subsection{Recovery of the Rouse and Prandtl equations}
\begin{figure} [ht]
\centerline{\includegraphics[angle=0,width=0.87\linewidth]{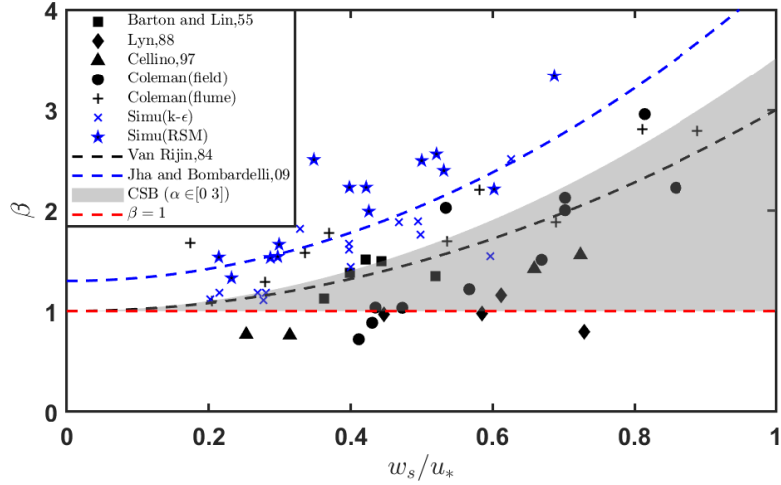}}
\caption{The model coefficient $\beta$ based on different formulae, experiments, and model runs. The experiments and model runs presented here are described elsewhere \cite{jha2009two}.}
\label{fig:alpha_beta}
\end{figure}

Whether a Rouse equation can be recovered from the CSB model under certain simplifications is now examined. Any explicit model must include $Sc$ and approximations to equation \ref{eq:sc}. Assuming $k_c/k_o \rightarrow 0$ and $k_o/k_e \rightarrow 0$ in $\Omega (z_n)$ only (i.e. setting the area under the Saffman spectrum to zero that is then partially compensated for by extending the inertial subrange to $k_e\rightarrow\infty$), the Schmidt number derived from equation \ref{eq:sc} can be approximated as
\begin{linenomath*}
\begin{equation} 
Sc^{-1} \approx 
1+B_{\pi}\left(\frac{w_s}{u_*}\right)^2, 
~\textrm{with}~ 
B_{\pi}= \frac{\sqrt{15A_{\pi}}}{3C_o}\alpha \approx 0.84 \alpha,
\label{eq:wc2}  
\end{equation}
\end{linenomath*}
which directly recovers the quadratic model for $Sc^{-1}$ reported elsewhere \cite{rijn1984sediment, bombardelli2012exchange} as expected. With $R \approx 0$, equation \ref{eq:wc2} indicate $\beta=Sc^{-1}\rightarrow 1$ thereby recovering Rouse's original assumption (i.e. SS resemble passive scalars in this case).  This estimate of $\beta$ also allows for the determination of the model coefficient $\alpha$ using a separate data set and model runs shown in Figure \ref{fig:alpha_beta}.

Figure \ref{fig:alpha_beta} shows different predictions of  $\beta$, including $\beta=1+2(w_s/u_*)^2$ \cite{rijn1984sediment} and $\beta=1.3+3(w_s/u_*)^2$ \cite{jha2009two} for model results that explicitly consider particle-fluid interactions. Moreover, with  $Sc$ provided in equation \ref{eq:wc2}, the SS diffusivity is derived as,
\begin{linenomath*}
\begin{equation} 
\frac{D_s(z)}{\kappa z u_*}=\frac{1}{Sc}\left(1-z_n \right)=\left[1
+B_{\pi}\left(\frac{w_s}{u_*}\right)^2 \right] \left(1-z_n \right)
\label{eq:mg}  
\end{equation}
\end{linenomath*}
where $\kappa z u_*$ is the eddy viscosity in the log-region of $\overline{u}(z)$. Depending on choices made for $\alpha$ or $B_{\pi}$, a number of empirical relations can be recovered including the widely used Rouse's equation and variants on it \cite{hunt1954turbulent}.  For a given $\alpha$, an analytical solution for the SSC can be derived and compared with published experiments.  The SSC solution for an arbitrary $\alpha$ is given as
\begin{linenomath*}
\begin{equation} 
\frac{\overline{C}(z_n)}{\overline{C_b}}
=\left(\frac{z_n}{1-z_n}\frac{1-z_{n,b}}{z_{n,b}}\right)^{-R_+}.
\label{eq:gs1}  
\end{equation}
\end{linenomath*}
where the power exponent $R_+$ is defined as
\begin{linenomath*}
\begin{equation} 
R_+=\frac{1}{1
+B_{\pi}\left({w_s}/{u_*}\right)^2  }
{\frac{w_s}{\kappa u_*}}.
\label{eq:rf}  
\end{equation}
\end{linenomath*}
When $\alpha=0$ (or $B_{\pi}=0$), a quadratic diffusivity profile \cite{obrien1933review} as well as Rouse's formula \cite{rouse1939analysis,rouse1937modern} for SSC given in equation \ref{eq:geS} are recovered. Furthermore, in the limit of ($z_n \ll 1$) a linear diffusivity profile \cite{vonKarman1934} along with the classic power law solution are also recovered from equation \ref{eq:gs1}. The consequences on $\sigma_w^2$ of setting the Saffman spectrum to zero and extending the inertial subrange to $k\rightarrow\infty$ on $\sigma_w^2$ are briefly discussed using Figure \ref{fig:etke}. As expected, these approximation over-estimate $(\sigma_w/u_*)^2$ in the near-wall region and underestimate $(\sigma_w/u_*)^2$ in the outer layer when compared to a $E_{ww}(k)$ that accommodates the Saffman spectrum (i.e. large scale effects) but truncates the inertial subrange at $1/k_e$.  These effects cannot be readily ignored and may influence the choices made about $\alpha$.

\subsection{Comparison with Experiments}
\begin{figure} [ht]
\centerline{\includegraphics[angle=0,width=0.99\linewidth]{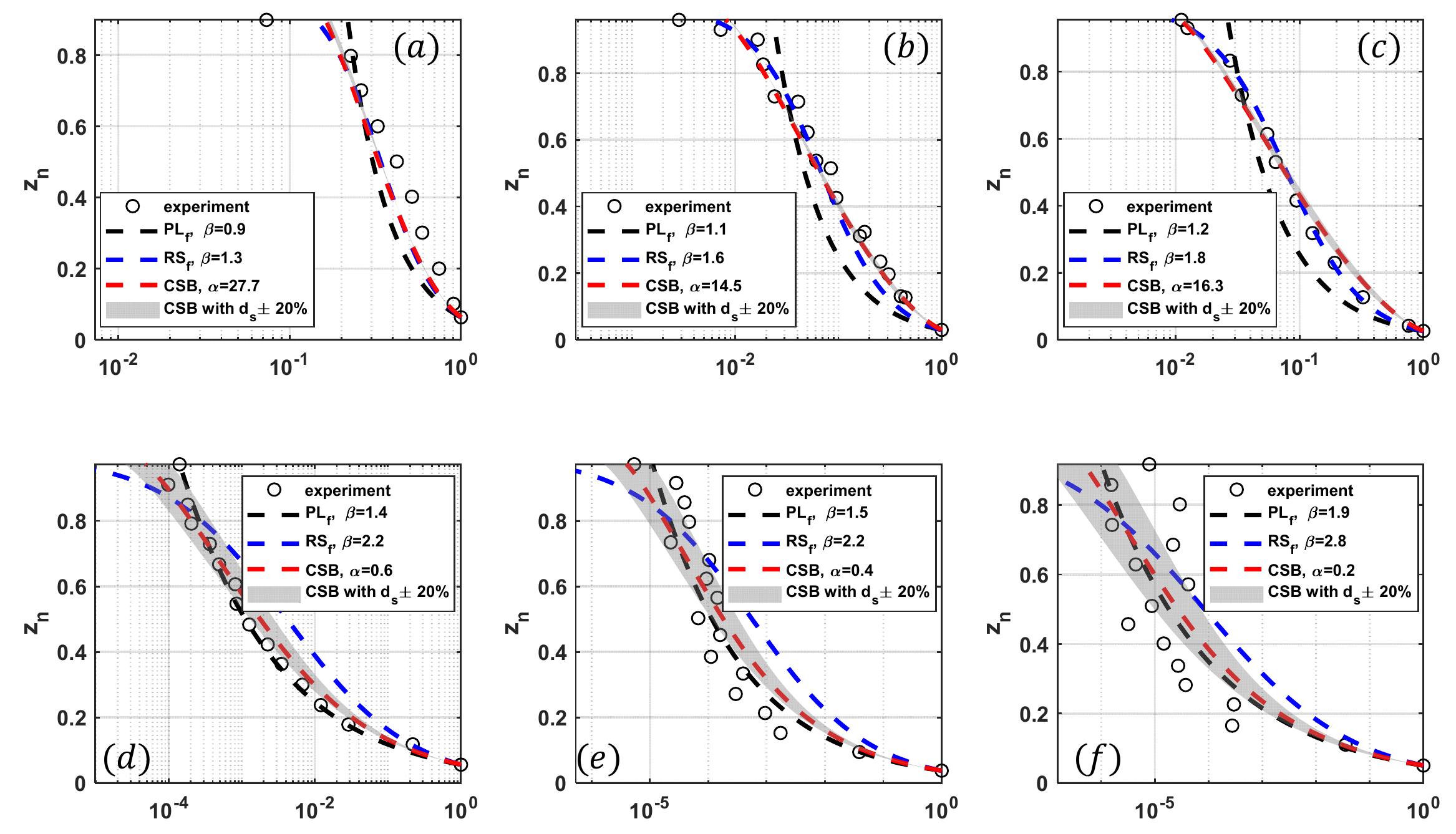}}
\caption{The predicted SSC profiles normalized by $C_b$ selected at the measurement height with the highest reported concentration. The panel labeling follows Table \ref {tb:tt1} with the top panel showing the comparisons from earlier measurement i.e. (a) \cite{xingkui1992velocity, greimann2001two} and (b)-(c) \cite{vanoni1984fifty}, and the bottom panel showing the comparisons using recent experiments (d)-(f) \cite{tsengtwo}.  For experiments in panels (a)-(c), the $St_b$ is not small ($>0.5$).}
\label{fig:SSC_comp}
\end{figure}

The CSB model given by equations \ref{eq:sscb1} and \ref{eq:sscb3} and its simplified version featured in equation \ref{eq:gs1} are compared with published experiments \cite{greimann2001two, vanoni1984fifty, tsengtwo} summarized in Table \ref{tb:tt1}. We assume $\Phi(z)=0$ thereby neglecting inertial effects for compatibility with operational models (e.g. the Rouse model).  The comparisons are shown in Figure \ref{fig:SSC_comp}. For these experiments, all the reported parameters including measured $u_*$, $d_s$, and $\rho_s$ and the fitted $\beta$ (needed for assessing the fitted Rouse equation) and $\alpha$ (needed for evaluating the numerical CSB model) are presented in Table \ref{tb:tt1}. 

\begin{table} 
\caption{Summary of published experiments and parameters used in model-data comparisons.  When setting $Sc$ =$1$, not all runs are classified as SS (or $0.8$ $\leq$ ${R^*}$ $\leq$ $2.5$) even though sediments were reported as suspended. While $St_b$ is not very small for (b) and (c), $St_+$ $\ll$ $1$. Calculated densimetric and critical Froude numbers ($Fr_d$ and $Fr_{dc}$) are presented along with the roughness Reynolds number $Re_{pa}$.  All experiments lie in the fully-rough ($Re_{pa}>100$) or transitional ($3<Re_{pa}<100$) regimes in fully-developed turbulence ($Re_b \geq 500$).}
\centering 
\begin{tabular}{lcccccc}
\hline
\textbf{Run}     & \textbf{(a)}& \textbf{(b)} & \textbf{(c)} & \textbf{(d)} & \textbf{(e)} & \textbf{(f)} \\ \hline
& \multicolumn{6}{c} {\textbf{Flow Properties}} \\
\hline
$H$ (m)          & 0.10           & 0.52           & 0.50           & 0.10           & 0.10           & 0.10           \\
$B$ (m)          & $-$           & 0.84           & 0.84           & 0.15           & 0.15           & 0.15           \\
$U_b$ (measured, m s$^{-1}$)      & 1.98           & 3.95           & 3.63           & 0.31           & 0.22           & 0.17           \\
$Re_b=U_b H/\nu  \times 10^{-4}$           & 18.7        & 193        & 170        & 2.9        & 2.1        & 1.6        \\
$Re_{pa}=u_*d_s/\nu$   &103	&169	&166	&16	&13	&7 \\
$u_*$ (cm s$^{-1}$)      & 7.67           & 20.0          & 20.0          & 1.7           & 1.4           & 0.8           \\ 
$z_{n,b} \times 10^{2}$ (measured)      & 6.3           & 2.6         & 3.0         & 5.0           & 3.8           & 5.6           \\ 
$z^+_{n,b}(= u_*z_{b}/\nu)$      & 456         & 2747       & 2988        & 85 
& 53           &45         \\ 
$U_b/u_*$ (measured)      & 25.8           & 19.4           & 18.2           & 18.2          & 15.7           & 21.3           \\
\hline
& \multicolumn{6}{c} {\textbf{Sediment Properties}} \\ 
\hline
$\rho_s/\rho$ & 1.05           & 2.65           & 2.65           & 1.20           & 1.20           & 1.20           \\
$d_s$ (mm)       & 1.42           & 0.88           & 0.88           & 1.00           & 1.00           & 1.00           \\
$w_s$ (cm s$^{-1}$)      & 1.7          & 10          & 10          & 2.9          & 2.9          & 2.9          \\
\hline
& \multicolumn{6}{c} {\textbf{Dimensionless Model Parameters}} \\
\hline
$R*=w_s/(\kappa u_*)$          & 0.5           & 1.2           & 1.2          & 4.3           & 5.2           & 9.1          \\
$\alpha$         &27.7	&14.5	&16.3	&0.6	&0.4	&0.2
           \\
$\beta$ (Rouse)         & 1.3           &1.6           &1.8         & 2.2           & 2.2           & 2.8           \\

$R=w_s/(\beta\kappa u_*)$          & 0.4           & 0.8           & 0.7          &1.9          & 2.4         & 3.2           \\

$\beta$  (Prandtl)        & 0.9          & 1.1          & 1.2           & 1.4           & 1.5           & 1.9           \\

$R=w_s/(\beta\kappa u_*)$          & 0.6           & 1.1           & 1.0         &3.0          & 3.5          & 4.8           \\

$St_b\left(={w_s}/{g} \sqrt{{g S_o U_b}/{\nu}}\right)$           & 0.57          & 5.63         & 5.4          & 0.09           & 0.06          & 0.03       
\\ 
$St_+\left(=\tau_pu_*/H\right) \times 10^{3}$           & 1.3          & 4.0          & 4.1        &0.5           & 0.4          & 0.2    \\
$Fr_d\left(=U_b/\sqrt{(\rho_s/\rho-1)gd}\right)$           & 75          & 33          & 30       &7        & 5          & 4 \\
$Fr_{dc}$           & 3          & 4          & 4       &3        & 3         & 3 \\
$U_b/u_*$ (CSB rough bed)    &12.4	&15.2	&15.6	&13.0	&14.0	&13.5 \\
$U_b/u_*$ (CSB smooth bed)    &27.2	&33.1	&33.5	&23.5	&23.7	&22.4 
\\ \hline  
\end{tabular}
\label{tb:tt1}
\end{table}
In the experiments, the sediments covered the bed and were assumed to have reached an equilibrium state where equation \ref{eq:fgov} applies \cite{tsengtwo}.  The densimetric Froude number $Fr_d$ and the critical densimetric Froude number $Fr_{dc}$ whose formulation is described elsewhere \cite{ali2017origin,ali2018impact,li2019cospectral} are also presented in Table \ref{tb:tt1}.  In all cases, the $U_b$, $\rho_s/\rho$, and $d/H$ result in $Fr_d>Fr_{dc}$ meaning that sediments can be released from the bed and must be balanced by sediments depositing onto the bed.  Thus, the experiments do not strictly abide by HAE's definition of SS as sediments here are not remain permanently suspended.  Across the experiments, the flow variables $U_b$ and $u_*$ varied from 10 cm s$^{-1}$ to 40 cm s$^{-1}$ and 0.8 to 8 cm s$^{-1}$, respectively.  However, $U_b/u_*=(8/f_{dw})^{1/2}$, related to the Darcy-Weisbach friction factor $f_{dw}$, varied much less (15-25) as may be anticipated in fully rough flow over a channel bed covered by grains of similar $d_s$.  The particle properties $\rho_s/\rho$ and $d_s$ varied from 1.05 to 2.65 and 0.88 to 1.4 mm, respectively.  The consequence of these variations is that the empirically derived settling velocity $w_s$ is much smaller than the Stokes settling velocity as shown in the inset of Figure \ref{fig:ws}. Collectively, these experiments span wide-ranging particle sizes (in the SS range) and flow properties from different sources.  The lowest measured sediment concentration near the channel bottom is close to the surface ($z_{n,b}\in [0.026, 0.063]$) but remains above the buffer region $z^+=u_*z_{b}/\nu>30$ as shown in Table \ref{tb:tt1}. For some runs, the $z^+<100$ and wall-blockage effects (not considered here) can impact $E_{ww}(k)$ and $d\overline{u}(z)/dz$ \cite{MccollEA16}, which introduce obvious uncertainties. As shown in Table \ref{tb:tt1}, experiments (a)-(c) are characterized by $St_b>0.5$, which may be indicative that $\Phi(z)$ is not small. Experiments (d)-(f) are characterized by a small $St_b$ as assumed by the CSB and Rouse's formula.

Figure \ref{fig:SSC_comp} confirms that the fitted Rouse formula and fitted Prandtl formula (i.e. the $R$ model, allowing $\beta$ to be fitted) offer good agreements with some measurements (for (a)-(b) and (d)-(f) respectively) at all depths. Given that the simplified CSB model is identical to Rouse's formula, an agreement between the fitted Rouses's formula and the measurements can also be juxtaposed to the simplified CSB model. However, the numerical CSB model provides reasonable agreements for all the runs when allowing $\alpha$ to vary.  Allowing $\alpha$ to be a free parameter has several advantages when compared to $\beta$ in the fitted Rouse equation. Setting $\beta$ as constant implies $Sc$ is constant at all $z_n$ while setting $\alpha$ as constant incorporates some of the local variations in $Sc$ with $z_n$ (albeit near the free water surface, maintaining a finite $\epsilon$ can be problematic without adjustments). The impact of minor variations in particle sizes is shown in the shaded area: the particle sizes are increased/decreased by 20\% to illustrate model sensitivity to $d_s$. Uncertainty in sediment composition (and thus $d_s$ and $w_s$) can be a factor in determining SSC uncertainty but not in all cases (runs d,e,f). 
While the SSC model does not require $\overline{u}$ (only $d\overline{u}/dz$), the predicted $U_b$ from the CSB turbulent stress budget can be compared against measured $U_b$ for a plausibility check. The modeled $U_b$ requires $u_*$ along with a boundary condition specified here as $\overline{u}(z_{n,b})/u_*$ at $z_{n,b}$.  A number of choices can be made about this boundary condition.  Given that $z_{n,b}$ is sufficiently distant from the wall, the most direct of those choices is the log-law for two end-member cases: (i) fully rough with an externally imposed surface roughness and (ii) hydrodynamically smooth.  In both cases, the mean velocity at $z_{n,b}$ is approximated as 
\begin{linenomath*}
\begin{equation} 
\frac{\overline{u}(z_b)}{u_*}=\frac{1}{\kappa} \log\left(\frac{z_b}{z_o} \right); \quad \quad \frac{\overline{u}(z_b)}{u_*}=\frac{1}{\kappa} \ln\left(z^+_{n,b}\right)+5,
\label{eq:loglaw_BC}  
\end{equation}
\end{linenomath*}
where $z_o$ is the momentum roughness length. The $z_o$ can be related to $d_s$ by $z_o \approx d_s/30$ where the grain diameter is assumed constant.  In all cases, the roughness Reynolds number $Re_{pa}=u_* d_s/\nu>3$ but in some cases, the flow is not fully rough (i.e. transitional with $3<Re_{pa}<100$).  For this reason, the CSB model forced by both rough and smooth surface boundary conditions at $z_{n,b}$ are featured in Table \ref{tb:tt1}.  The agreement between measured and the range of CSB modeled $U_b/u_*$ for these two end-member cases appears reasonable. Runs (a) and (f) are closer to a smooth-wall case whereas runs (b), (c), and (e) are better approximated by a rough-wall boundary condition. Run (d) falls in-between these two end-member cases.  While Run (f) had the smallest $Re_{pa}=8$ and a near-smooth wall approximation may be justifiable, run (a) had an $Re_{pa}>100$. We do not have a clear explanation as to why $U_b$ in run (a) is better approximated by a smooth wall boundary condition.

\begin{figure} [ht]
\centerline{\includegraphics[angle=0,width=0.63\linewidth]{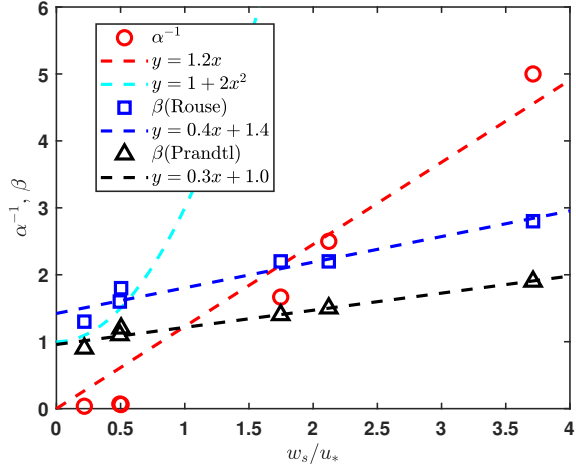}} 
\caption{The dependence of fitted $\alpha$ and $\beta$ on the $w_s/u_*$. The red, blue and black dashed lines show the fitted trend-lines of $\alpha^{-1}$ and $\beta$ from Rouse and Prandtl equations respectively.  The cyan dashed line is $\beta$=$1+$ $2(w_s/u_*)^2$ \cite{rijn1984sediment} extrapolated for large $w_s/u_*$.}
\label{fig:Sc_beta}
\end{figure}

An investigation of the relation between fitted $\alpha$ (and $\beta$) and $w_s/u_*$ is undertaken and shown in Figure \ref{fig:Sc_beta}.  A near-linear relation between $\alpha^{-1}$ and $w_s/u_*$ indirectly supports the heuristic closure adopted for $\overline{C'{\partial w'}/{\partial z}}$ with some caveats.   

In the regime $w_s/u_*\gg 1$, the closure model with $b_1\sim \mathrm{sgn}(A_f) u_*/w_s$ leads to an $\alpha^{-1}\sim -\mathrm{sgn}(A_f)(1-C_I)(w_s/u_*)$ and $\beta\sim -\mathrm{sgn}(A_f)/(1-C_I)(w_s/u_*)$, both of which are negative unless $\mathrm{sgn}(A_f)$ is negative. The relation in Figure \ref{fig:Sc_beta} indicates a positive slope between fitted $\alpha^{-1}$ and $w_s/u_*$, suggesting that the coefficient $A_f$ in the flux-variance similarity closure (i.e. equation \ref{eq:stressfluxbudget4}) is negative.  More broadly, to what extend this closure is general and how robust are its results in the context of SSC profile predictions cannot be unpacked from the experiments here and is better kept for a future research topic.  

\section{Model Limitations}
The treatment of suspended sediments as a dilute mixture is an obvious model limitation. This assumption requires particles to settle independently and that the solid volume can be ignored relative to the water volume. For the experiments considered here, this assumption is reasonable. Another restrictive assumption is setting $\Phi=0$ \cite{kind1992one,chamecki2007concentration}. A $\Phi=0$ also leads to $\overline{C}=\overline{w'C'}/w_s\rightarrow0$ at $z_n\rightarrow1$, which may not be general. Given the large vertical gradients in $\sigma_w^2$ near the channel bottom and near the free water surface, turbophoretic effects can be significant in these two regions \cite{caporaloni1975transfer,guha1997unified,marchioli2002mechanisms,zhao2006modeling,katul2010predicting,chamecki2007concentration}. The turbophoretic effect act to increase the SS concentration near the water surface; however, the measurements here (runs a-c) suggest that for the $St_b>1$ cases, the SS concentrations near the water surface experience a decline as $z_n\rightarrow1$ instead of an increase. This finding can be used to suggest that $\Phi=0$ may be plausible as the turbophoretic term was shown to dominate $Phi$ near the water surface \cite{richter2018inertial,bragg2021mechanisms}. The CSB budget formulation here (i.e. equation \ref{eq:vscg}) ignored the flux transfer term and their vertical variation.  In the case of the turbulent stress, ignoring the flux transfer term (and its vertical gradients) altogether guarantees that the co-spectrum between $w'$ and $u'$ in the inertial subrange maintains a $k^{-7/3}$ scaling.  This $k^{-7/3}$ scaling has been observed in numerous boundary layer studies reporting co-spectra thereby offering indirect justification for this assumption. The flux transport terms (i.e. the vertical gradients of triple moments in the Reynolds averaged equations) have also been ignored.  These terms have been studied less for stress and sediment flux turbulent budgets compared to their turbulent kinetic energy budget counterparts.  The work here highlights the need for an assessment of these terms relative to their mechanical production terms.  The CSB model also assumes that the linear Rotta scheme (slow component) with an isotropization of production (rapid component) applies equally to SS and momentum fluxes without adjustments in constants (i.e. $A_R=1.8$ and $C_I=3/5$).  Hence, any departure from these established constants must be absorbed by $t_{ww}(k)/t_r(k)$, which manifests itself as a Schmidt number effect (or $\alpha$ variations).

The assumed shape of $E_{ww}(k)$ is also over-simplified and certainly not reflective of what is known about the energetics near the surface ($z^+<100$) such as wall-blockage.  Moving away from the wall region itself, other 'shape issues' arise.  For example, near the spectral transition from inertial to viscous regimes, usually occurring at around $k\eta \approx 0.1$, $E_{ww}(k)$ experiences a bottleneck that is absent here \cite{saddoughi1994local,katul2015bottlenecks}.  Likewise, as $k\eta>0.1$ and increases further into the viscous regime, $E_{ww}(k)$ decays exponentially \cite{pope2001turbulent}.  Hence, extending the inertial subrange to $k\eta=1$ is not intended to capture all such mechanisms impacting the vertical velocity spectrum.  Instead, it allows for some compensation of loss in energy due to censoring $E_{ww}(k)$ at $k\eta =1$ while introducing extra energy due to an expected overestimation of the extrapolated inertial subrange spectrum in this vicinity. On a more positive note, while the full details of the turbulent kinetic energy cascade across scales are not explicitly considered, their effects remain implicitly contained in the assumed shape of $E_{ww}(k)$.  As such, some of these effects can be accommodated (e.g. the bottleneck, viscous cutoff, etc...) by various revisions to $E_{ww}(k)$ (e.g. including a bump around $k\eta =0.1$, resolving the viscous cutoff region using the Pao spectral shape or variants \cite{pope2001turbulent} on it, etc...). 

It is to be noted that the co-spectral budget is integrated scale-wise, which means that the precise shape of $E_{ww}(k)$ in the vicinity of $k\eta \approx 1$ is less crucial.  Moving beyond the shape issues of $E_{ww}(k)$ and focusing on its primary input variable $\epsilon(z_n)$, the approach assumes turbulent kinetic energy production is balanced by its dissipation at every $z_n$ (i.e. $\phi(z_n)=1$), which is certainly not realistic for all $z_n$. However, as previously mentioned, deviations from unity in $\phi(z_n)$ may be ameliorated by the sub-unity exponent ($-1/3$) dependence in the SSC budget. An exception to this statement is the particle time scale $t_{wc}(k)$ in $Sc(k)$.  A $\phi(z_n)=1$ as $z_n\rightarrow1$ leads to an unbounded $Sc^{-1}(k)$ and thus an uncertain $D_s$ shape in the vicinity of the free surface. A plausible adjustment to the $Sc^{-1}(k)$ calculations based on maintaining a minimal $\epsilon$ ($=0.001\epsilon_b$) was introduced here though this correction remains adhoc.  Last, the turbulent SS flux from the CSB model(s) follows the same form as gradient-diffusion closure upon ignoring both - turbulent flux transport and scale-wise transfer terms. However, a key advantage here is that the effective diffusion coefficient $D_s$ from the CSB model contains contributions from turbulent eddies and Schmidt numbers at all scales. The proposed Schmidt number (or $\alpha$) is consistent with bulk Schmidt number formulations such as those by van Rijin's and other one-way coupling schemes (i.e. particle transport does not impact the flow) when $Sc<1$ \cite{bombardelli2012exchange}. For dense mixture or other aeolian particles in the atmosphere, the particle Schmidt number can be larger than unity \cite{csanady1963turbulent} implying other particle-fluid interaction models are required. 
When using the CSB model, the $\alpha$ used for the determination of the Schmidt number is treated as a single fitted parameter.  Hence, the CSB model offers the same number of free parameters as the fitted Rouse equation. What was found here is that $\alpha^{-1}$ varies linearly with $w_s/u_*$ when combining all the experiments.  A plausibility argument as to why $\alpha$ depends on $w_s/u_*$ was also offered.  In some instances, the addition of a single fitted parameter may be desirable in hydraulic models as discussed elsewhere \cite{papke2013reduced, battiato2014single, rubol2018universal, li2019mean}, but an increasing number of free model parameters does not necessarily lead to a better physical understanding. The sediment settling velocity estimated in equation \ref{eq:ws} is commonly based on a mass-median-diameter from particle size distribution measurements, which however may not be an optimized characteristic size as shown by some in-situ measurements \cite{williams2007high}.  Large variations in $d_s$ can have a substantial impact on SSC profiles, which may be more significant than models for $\alpha$.

\section{Conclusion}
Operational modeling of SSC in turbulent flows continues to be a formidable challenge in hydraulics, hydrology, ecology, and water quality control.  The work here establishes a new link between the spectrum of vertical velocity and SS turbulent flux, which was then used to arrive at expressions for the SSC profile.  The spectrum of vertical velocity is characterized by multiple scaling regimes that include the Saffman spectrum ($E_{ww}(k) \sim k^{+2}$), the 'energy splashing' effect due to the presence of a wall ($E_{ww}(k) \sim k^{0}$), and the much-studied inertial subrange regime ($E_{ww}(k) \sim k^{-5/3}$).  Finite Reynolds effects are accommodated through a scale separation between $z$ and the Kolmogorov microscale $\eta$ terminating the scale-wise extent of the inertial subrange (as a first approximation).  This dependence can be noted when considering the scaling argument $k_e/k_o = z/\eta \sim (z u_*/\nu)^{3/4}$ \cite{tennekes2018first}.  Hence, increasing $Re_s=(z u_*/\nu)$ by either increasing $z$ or $u_*$ leads to a widening of the scale-wise extent of the inertial subrange, which then impacts all subsequent expressions such as $\Omega(z_n)$ and $d\overline{u}/dz$.  As such, the proposed model is responsive to finite Reynolds number, Schmidt number, and Rouse number effects. Prior \textit{ad-hoc} efforts such as correcting $l_o$ by $V_n$ (i.e. the van Driest damping function) can now be interpreted from this new spectral perspective (i.e. $Re_s$ effects become large for small $z$ or $u_*$).  A simplified solution to the CSB model in which the Saffman spectrum is truncated but the inertial subrange is now extended to infinite wave-numbers (i.e. $Re_s\rightarrow\infty$) was shown to recover earlier theories (e.g. Rouse's formula).  The fitted Rouse's equation (and by extension the simplified CSB solution) also describes the measured SSC profiles in all the experiments considered here provided $\alpha$ (or $\beta$) is allowed to vary with $w_s/u_*$.  Thus, one of the main novelties here is to provide a spectral link between the energy distribution in eddies and the SSC shape. Interactions between turbulent eddies and suspended sediment grains at various heights were also proposed, resulting in a scale-dependent $Sc$ captured by a single parameter $\alpha$ that varies with $w_s/u_*$. Such $Sc$ variations were formulated in spectral space but recover expected bulk relations between $R$ and $Sc$ identified by other models, experiments, and simulation studies.  When all these findings are taken together, future extension of this work must focus on upgrading the particle-turbulence interaction scheme and its signature in a scale-dependent Schmidt number.  Such extension will benefit from targeted DNS runs where all the terms in the particle co-spectrum as well as $E_{ww}(k)$ can be computed or determined.  Likewise, an exploration of where the sediment flux transport term is significant relative to the mechanical production term and how to incorporate its effects can be undertaken from the aforementioned DNS runs.

\subsection*{Data Availability and Acknowledgements}
All the data used were digitized from the published literature \cite{greimann2001two, vanoni1984fifty, jha2009two, tsengtwo}. SL was supported by a fellowship from the Nicholas School of the Environment at Duke University.  GK and ADB acknowledge support from the U.S. National Science Foundation (NSF-AGS-1644382, NSF-AGS-2028633, NSF-IOS-1754893, and NSF-CBET-2042346). 

\bibliography{2_CSB-SSC.bib}
\end{document}